\documentclass[unsortedaddress,aps,preprint,showpacs]{revtex4}
\usepackage{graphicx}
\usepackage{dcolumn}
\usepackage{amsmath}
\usepackage{amssymb}
\usepackage{amsfonts}
\usepackage{bm}
\usepackage{color}
\usepackage[normalem]{ulem}
\newcommand{\be}{\begin{equation}}
\newcommand{\ee}{\end{equation}}
\newcommand{\ba}{\begin{eqnarray}}
\newcommand{\ea}{\end{eqnarray}}

\begin{document}
\title{Clusterization of weakly-interacting bosons in one dimension:\\An analytic study at zero temperature}
\author{Santi Prestipino$^1$\footnote{Corresponding author. Email: {\tt sprestipino@unime.it}}, Alessandro Sergi$^{1,2}$\footnote{Email: {\tt asergi@unime.it}}, and Ezio Bruno$^3$\footnote{Email: {\tt ebruno@unime.it}}}
\affiliation{$^1$Universit\`a degli Studi di Messina,\\Dipartimento di Scienze Matematiche e Informatiche, Scienze Fisiche e Scienze della Terra,\\viale F. Stagno d'Alcontres 31, 98166 Messina, Italy\\$^2$Institute of Systems Science, Durban University of Technology, P.\,O.\,Box 1334, Durban 4000, South Africa\\$^3$Universit\`a degli Studi di Messina,\\Dipartimento di Ingegneria,\\c.da Di Dio, 98166 S. Agata, Messina, Italy}
\date{\today}

\begin{abstract}
We study a system of penetrable bosons on a line, focusing on the high-density/weak-interaction regime, where the ground state is, to a good approximation, a condensate. Under compression, the system clusterizes at zero temperature, i.e., particles gather together in separate, equally populated bunches. We compare predictions from the Gross-Pitaevskii (GP) equation with those of two distinct variational approximations of the single-particle state, written as either a sum of Gaussians or the square root of it. Not only the wave functions in the three theories are similar, but also the phase-transition density is the same for all. In particular, clusterization occurs together with the softening of roton excitations in GP theory. Compared to the latter theory, Gaussian variational theory has the advantage that the mean-field energy functional is written in (almost) closed form, which enables us to extract the phase-transition and high-density behaviors in fully analytic terms. We also compute the superfluid fraction of the clustered system, uncovering its exact behavior close, as well as very far away from, the transition.
\end{abstract}

\maketitle
\section{Introduction}

The quantum mechanics of one-dimensional (1D) many-body systems is, since many decades by now, a well-respected area of research at the boundary between physics and mathematics, with lots of exact results~\cite{Cazalilla}. However, until recent times 1D systems were deemed to be of little interest for real-world physics. Now, their study is experiencing a revival thanks to the availability of ultracold gases of atoms and dipolar molecules confined in highly anisotropic magnetic traps and optical lattices (owing to strong confinement in the transverse directions, only the lowest-energy transverse quantum state needs to be considered, and the three-dimensional problem becomes effectively one-dimensional). Furthermore, by tuning the transverse trap frequency it is possible to change the short-range part of the atom-atom interaction to a certain extent (see, e.g., Ref.~\cite{Bergeman}), opening the way to a direct comparison between experiment and theory. 

Early experimental and simulation works have focused on reproducing the physics of the exactly solvable Lieb-Liniger model~\cite{Lieb1,Lieb2}, describing a 1D system of identical spinless bosons of mass $m$ interacting through a contact repulsion of strength $g$. For all values of the single parameter $\gamma=mg/(\hbar^2\rho)$ characterizing the model (with $\rho$ denoting the number density), the spectrum is phonon-like at small momenta, in accordance with Luttinger-liquid (LL) behavior~\cite{Haldane}. The system becomes more weakly-interacting with {\em increasing} $\rho$; indeed, it is when $\gamma$ is low that, according to a standard argument (see, e.g., \cite{Petrov}), the coherence (or healing) length $\hbar/\sqrt{mg\rho}$ ($m$ is the atom mass) is much larger that the mean interparticle separation $\rho^{-1}$. For small values of $\rho/g$ (large $\gamma$), the 1D fluid acquires fermionic properties as the ground-state wave function strongly decreases at short interparticle distances (Tonks-Girardeau limit)~\cite{Girardeau}. At zero temperature ($T=0$), the one-body density matrix exhibits a power-law decay at large distances for any $g$ (although with a $\rho$-dependent exponent becoming smaller and smaller as $\rho/g$ increases); therefore, strictly speaking there is no Bose-Einstein condensation. It was Hohenberg~\cite{Hohenberg} the first to rigorously prove that no form of long-range order (including off-diagonal long-range order) can exist in one- and two-dimensional quantum systems with continuous group symmetries for non-zero temperatures; later, Pitaevskii and Stringari~\cite{Pitaevskii2} have extended the proof for 1D bosonic systems also to the zero-temperature case. However, if we add a trapping potential in the axial direction both the amplitude and phase fluctuations are suppressed at low $T$ and one has a true condensate in 1D, provided the number $N$ of particles is large enough~\cite{Petrov}.

Kinoshita and coworkers~\cite{Kinoshita} have prepared a gas of ultracold $^{87}$Rb atoms in a 1D, cigar-shaped region. Acting on the harmonic confinement in the transverse directions, they have been able to tune the coupling strength $\gamma$ in the axial direction, making atoms to resemble either a Bose-Einstein condensate ($\gamma\ll 1$) or a Tonks-Girardeau gas of impenetrable bosons ($\gamma\gg 1$). In a wide range of $\gamma$, the experimental data of Kinoshita {\it et al.} fit the exact solution for the ground state of the Lieb-Liniger model. Astrakharchik and Giorgini~\cite{Astrakharchik} have studied the same model by Monte Carlo simulation at $T=0$, again observing the crossover from the low-density/free-fermion/Tonks-Girardeau limit to the high-density/weak-interaction/Gross-Pitaevskii (GP) regime~\cite{Gross1,Pitaevskii1,Gross2}.

In 1D systems without pinning or trapping potentials, the Tonks-Girardeau behavior is typically confined to low density; for dense systems other ground states may emerge, like in dipolar bosons, which form a crystal at very high density~\cite{Arkhipov}. A further example is a system of softly-repulsive bosons. In a classical context, it has been recognized that bounded potentials favor clustering at low temperature, i.e., crystals with multiply-occupied sites, even when the potential is purely repulsive~\cite{Likos,Mladek}. Similarly, quantum cluster crystals have been predicted in dense 2D systems of penetrable bosons~\cite{Pomeau,Josserand,Henkel,Saccani,Cinti,Macri1}. In 1D, thermal fluctuations destabilize cluster-crystalline order in favor of cluster-dominated liquid phases with different average occupation~\cite{Speranza,Prestipino1,Prestipino2}. Quantum mechanics induces coherent delocalization even at $T=0$~\cite{Pitaevskii2}, transforming a cluster crystal into a cluster LL (CLL)~\cite{Rossotti,Teruzzi1,Teruzzi2}. On decompression, the latter phase undergoes a quantum phase transition into a LL without clusters. Rossotti {\em et al.}~\cite{Rossotti} have studied in detail the phase transition from LL to ``dimer'' CLL at $T=0$, which turns out to fall into the Ising universality class.

We here reconsider the $T=0$ phase diagram of weakly-interacting penetrable bosons in 1D, but now using the GP theory for the ground state, namely assuming a pure condensate from the outset. While this cannot be valid in general, the mean-field (MF) approximation is reasonable in the high-density region, which comprises the transition from the fluid to a high-occupancy CLL. As we did for the same system in two and three dimensions~\cite{Prestipino3}, we describe the CLL phase using a variational wave function written as a sum of evenly spaced Gaussians~\cite{Tarazona}. Even though no truly long-range order can exist in an infinite 1D system at $T=0$~\cite{Pitaevskii2}, modeling the CLL phase as a crystal entails an error on the system energy and short-range spatial correlations that is small in the whole range of applicability of MF theory. Otherwise, we may think of a large system of particles arranged in a circle (i.e., a finite system with periodic conditions at its boundary); in this case, clusterization and crystallization become one and the same transition. We show that the optimal Gaussian condensate is an effective approximation to the exact GP ground state. Gaussian variational theory has a major advantage over GP theory, since it allows an analytic study of both the transition region and the ultra-high-density limit, further enabling us to compute the superfluid fraction of the cluster phase.

Before going any further, it is worth discussing the relevance of MF approximation for the effectively 1D systems that can be realized experimentally. For atoms in a cylindrical trap, with transverse level spacing $\hbar\omega_\perp$ greatly exceeding the MF energy $g\rho$, the radial extension of the wave function is $a_\perp=\sqrt{\hbar/(m\omega_\perp)}$. As discussed e.g. in \cite{Pitaevskii3}, radial motion is frozen only when the product of the 1D density $\rho$ times the 3D scattering length $a_{\rm 3D}$ is very small ($\rho a_{\rm 3D}\ll 1$). On the other hand, the MF regime in 1D corresponds to $\rho|a_{\rm 1D}|\gg 1$ ($a_{\rm 1D}=-2\hbar^2/(mg)$ being the 1D scattering length), which is consistent with $\rho a_{\rm 3D}\ll 1$ only provided that $a_{\rm 3D}\ll a_\perp$ (since $a_{\rm 1D}=-a_\perp^2/a_{\rm 3D}$ for values of $a_{\rm 3D}$ far below the threshold $a_\perp/1.0326$ of confinement-induced resonance~\cite{Olshanii}). The smaller the ratio $a_{\rm 3D}/a_\perp$ is, the wider the range of densities where the MF approximation holds. In principle, this range can be expanded by reducing the soft-core interaction strength --- which, however, might be hard to achieve in practice.

The outline of the paper is the following. In Sec.\,II we introduce the model and the theory employed to study its phase behavior at $T=0$. In Sec.\,III we present our results. Concluding remarks are postponed to Sec.\,IV.

\section{Model and theory}
\setcounter{equation}{0}
\renewcommand{\theequation}{2.\arabic{equation}}

We investigate a system of $N$ one-dimensional (1D) spinless bosons of mass $m$, interacting through a bounded potential $u(x)$, even function of its argument. The range $\sigma$ and strength $\epsilon$ of the potential set the units of length and energy, respectively. A paradigmatic example of bounded repulsion is the penetrable-sphere model (PSM) potential, $u(x)=\epsilon\Theta(\sigma-|x|)$, $\Theta$ being the Heaviside step function. In the MF approximation, the ground state of the system is represented as a pure condensate:
\be
\Psi(x_1,\ldots,x_N)=\prod_{i=1}^N\psi(x_i)\,.
\label{eq2-1}
\ee
The best choice of $\psi$ is that minimizing the expectation value of the Hamiltonian in the state $\Psi$, which corresponds to a single-particle wave function obeying the (time-independent) Gross-Pitaevskii (GP) equation (see, e.g., Ref.\,\cite{Rogel-Salazar}):
\be
-\frac{\hbar^2}{2m}\psi^{\prime\prime}(x)+N\int_V{\rm d}y\,|\psi(y)|^2u(x-y)\psi(x)=\mu\psi(x)\,,
\label{eq2-2}
\ee
where $\mu$ has to be adjusted so that $\psi$ is normalized:
\be
\int_V{\rm d}x\,|\psi(x)|^2=1\,.
\label{eq2-3}
\ee
Hereafter, the 1D volume $V$ is considered as macroscopic, with $\rho=N/V$ finite (wherever appropriate, an integral over $V$ can be extended to the whole real axis).

In the aim to describe clusterization of the system at $T=0$, we make the ansatz
\be
\psi(x)=\frac{1}{\sqrt{V}}\sum_Gc_Ge^{iGx}\,,
\label{eq2-4}
\ee
where $G=(2\pi/a)n$ are reciprocal-lattice vectors and $\sum_G|c_G|^2=1$. The state function in Eq.\,(\ref{eq2-4}) describes a 1D crystal of spacing $a$. As discussed in the Introduction, no long-range order can actually occur in 1D, even for $T=0$; pretending the opposite is true is clearly an approximation, which predicts the wrong decay of correlation functions at large distances but only barely affects the location of the clusterization transition in the $\epsilon\rightarrow 0$ limit.

Plugging Eq.\,(\ref{eq2-4}) in the GP equation, we obtain~\cite{Kunimi,Prestipino3}
\be
\left(\frac{\hbar^2K^2}{2m}+\rho\widetilde{u}(0)\right)c_K+\rho\sum_{G\ne 0}\widetilde{u}(G)S_Gc_{K+G}=\mu c_K\,,
\label{eq2-5}
\ee
with $S_G=\sum_{G'}c_{G'}c_{G'+G}^*$. The function $\widetilde{u}(k)$ is the real-valued Fourier transform of $u(x)$, satisfying $\widetilde{u}(k)=\widetilde{u}(-k)$. The fluid phase, corresponding to $c_G=\delta_{G,0}$, is a special solution to Eq.\,(\ref{eq2-5}), with $\mu=\rho\widetilde{u}(0)$. For the $\psi(x)$ in Eq.\,(\ref{eq2-4}), the MF energy per particle is given by~\cite{Kunimi,Prestipino3}
\ba
{\cal E}&=&-\frac{\hbar^2}{2m}\int_V{\rm d}x\,\psi^*(x)\psi''(x)+\frac{N}{2}\int_V{\rm d}x\int_V{\rm d}x'\,|\psi(x')|^2u(x-x')|\psi(x)|^2
\nonumber \\
&=&\frac{\hbar^2}{2m}\sum_GG^2|c_G|^2+\frac{\rho}{2}\sum_{G_1,G_2,G_3}\widetilde{u}(G_1)c_{G_1+G_2}^*c_{G_1+G_3}c_{G_2}c_{G_3}^*\,.
\label{eq2-6}
\ea
The fluid energy is $\rho\widetilde{u}(0)/2$. We see from Eq.\,(\ref{eq2-6}) that, denoting $e_0$ the characteristic energy $\hbar^2/(m\sigma^2)$, the system ground state is only controlled by the dimensionless quantity $\rho\sigma\epsilon/e_0$ (which in the following is referred to as the ``density'') or, equivalently, by $\rho\widetilde{u}(0)/e_0$. A yet different expression of ${\cal E}$ is obtained in the Madelung representation, where the single-particle wave function is written as
\be
\psi(x)=\frac{1}{\sqrt{V}}\eta(x)e^{i\theta(x)}
\label{eq2-7}
\ee
(the amplitude $\eta(x)$ and phase $\theta(x)$ of $\psi$ are real and periodic). One readily obtains~\cite{Pomeau}:
\be
{\cal E}=\frac{\hbar^2}{8mV}\int_V{\rm d}x\left(\frac{\eta^{\prime 2}(x)}{\eta(x)}+4\eta(x)\theta^{\prime 2}(x)\right)+\frac{\rho}{2V}\int_V{\rm d}x\int_V{\rm d}x'\,\eta(x')u(x-x')\eta(x)\,,
\label{eq2-8}
\ee
which makes it clear that the ground-state wave function is necessarily real.

It is useful to discuss the range of applicability of MF theory as a function of system dimensionality $d$. As mentioned in the Introduction, MF theory is expected to hold when the healing length $\hbar/\sqrt{mg\rho}$ (with $g=\widetilde{u}(0)\approx\sigma^d\epsilon$), fixing the length scale above which collective physics dominates over single-particle physics~\cite{Pitaevskii3}, is much larger than the average interparticle separation $\rho^{-1/d}$~\cite{note}. In 3D, this leads to $\rho\sigma^3\epsilon/e_0\ll(e_0/\epsilon)^2$, which corresponds to a density range that is wider the weaker the interaction strength. However, in 1D the MF regime is rather $\rho\sigma\epsilon/e_0\gg(\epsilon/e_0)^2$, and the approximation improves with increasing density.

The way to solve Eqs.\,(\ref{eq2-5}) for fixed values of $\rho$ and $a$ is by iteration~\cite{Kunimi}: at each step of the procedure, $S_G$ is first estimated from the $c_G$ coefficients computed at the previous step; the resulting linear system is then solved, determining eigenvalues $\mu_n$ and normalized eigenvectors. Next, the coefficients are updated to the eigenvector with minimum energy. The iterative process comes to an end when self-consistency is attained. The final task to accomplish is the optimization of the lattice parameter $a$, which is stopped when its value is determined to six decimal places. Once the specific energy $e$ has been computed as a function of $\rho$, the identification of the stable ground state at pressure $P$ proceeds via the minimization of the generalized enthalpy $\widetilde{h}(\rho;T=0,P)=e(\rho)+P/\rho$, which contextually determines the equilibrium density as $\rho_{\rm eq}(P)={\rm argmin}\,\widetilde{h}(\rho)$.

Kunimi and Kato have solved Eqs.\,(\ref{eq2-5}) for PSM bosons in two dimensions (2D)~\cite{Kunimi}, showing that the high-density ground state is a triangular crystal. Macr\`i {\em et al.}~\cite{Macri2} have tested MF results by Monte Carlo simulation, finding that the condensate is indeed only weakly depleted in the fluid region and that the exact freezing point is close to the theoretical estimate. In Ref.~\cite{Prestipino3} we have extended the ground-state calculation to other 2D and 3D lattices, by employing an accurate variational form of $\psi$ --- written as a sum of Gaussians centered at the lattice sites --- that reproduces MF data to a high degree of accuracy. By this method, we have shown that the $T=0$ phase diagram of selected 3D potentials can be reconstructed with modest computational effort. Here, we make the same ansatz on the shape of the 1D wave function. While performing well in comparison with unconstrained MF theory, Gaussian variational theory has the distinct virtue of allowing a number of analytic shortcuts that considerably simplify extracting physical predictions from MF theory.

Using the variational method, we represent the single-particle state by the real-valued wave function
\be
{\rm VT}1:\,\,\,\,\,\,\psi(x)=C_\alpha\frac{1}{\sqrt{V}}\sum_{n=-\infty}^{+\infty}e^{-\alpha\left(\frac{x}{a}-n\right)^2}\,,
\label{eq2-9}
\ee
where $C_\alpha$ is a suitable normalization constant (observe that the fluid phase, where $\psi=1/\sqrt{V}$, is recovered as a special case of (\ref{eq2-9}), for $\alpha\rightarrow 0$). Two variational parameters are present in Eq.\,(\ref{eq2-9}), i.e., $\alpha$ and $a$, related to the width and periodicity of the Gaussians, respectively (we stress that $a$ is an adjustable quantity as well, so as to ensure the possibility of cluster-crystal states). The best parameters ($\overline{\alpha}$ and $\overline{a}$) are those minimizing the restriction ${\cal E}(\alpha,a;\rho)$ of functional (\ref{eq2-6}) to the set of functions (\ref{eq2-9}); once $\overline{\alpha}$ and $\overline{a}$ have been computed for each density, the energy per particle is given by $e(\rho)={\cal E}(\overline{\alpha}(\rho),\overline{a}(\rho);\rho)$ (there is an energy branch for the crystal and another, $e(\rho)=(\widetilde{u}(0)/2)\rho$, for the fluid). Denoting $N_c$ the number of crystal cells, the average number of particles in a cell is $N/N_c=(N/V)(V/N_c)=\rho a$. Hence, in a MF setting a cluster crystal is a crystalline state with $\rho a>1$. We denote VT1 the variational theory based on Eq.\,(\ref{eq2-9}); a different variational approximation, denoted VT2, will be considered below.

In explicit terms,
\be
C_\alpha=\left(\frac{2\alpha}{\pi I(\alpha)^2}\right)^{1/4}\,,\,\,\,\,\,\,{\rm with}\,\,\,\,\,\,I(\alpha)=\sum_{n=-\infty}^{+\infty}e^{-\frac{\alpha}{2}n^2}\,.
\label{eq2-10}
\ee
Although it does not admit an expression in terms of elementary functions, $I(\alpha)$ is related to a Jacobi theta function [cf. Eq.\,(\ref{a-1}) in the Appendix]:
\be
I(\alpha)=\vartheta_3(0,e^{-\alpha/2})\,.
\label{eq2-11}
\ee
The periodic function $\psi(x)$ can also be written as a Fourier series:
\be
\psi(x)=\frac{1}{\sqrt{V}}\sum_G\psi_Ge^{iGx}\,,
\label{eq2-12}
\ee
with coefficients~\cite{Prestipino3}
\be
\psi_G=C_\alpha'e^{-\frac{G^2a^2}{4\alpha}}\,\,\,\,\,\,{\rm and}\,\,\,\,\,\,C_\alpha'=\left(\frac{2\pi}{\alpha I(\alpha)^2}\right)^{1/4}\,.
\label{eq2-13}
\ee
From the normalization condition $\sum_G\psi_G^2=1$ we derive another expression for $I(\alpha)$, namely
\be
I(\alpha)=\left(\frac{2\pi}{\alpha}\right)^{1/2}\sum_Ge^{-\frac{G^2a^2}{2\alpha}}\,.
\label{eq2-14}
\ee
Equation (\ref{eq2-14}) proves useful to develop a low-$\alpha$ expansion of the energy functional (see Sec.\,III).

The advantage of the Gaussian series (\ref{eq2-9}) over the more general function (\ref{eq2-4}) is to allow analytic manipulations that considerably simplify the energy functional ${\cal E}(\alpha,a;\rho)$. Repeating the same steps followed in Ref.~\cite{Prestipino3}, we first obtain a closed-form expression for the zero-point kinetic energy:
\be
{\cal E}_{\rm kin}=e_0\frac{\alpha\sigma^2}{2a^2}\left(1+2\alpha\frac{I'(\alpha)}{I(\alpha)}\right)\,.
\label{eq2-15}
\ee
As for the potential-energy functional, it simplifies to~\cite{Prestipino3}:
\ba
{\cal E}_{\rm pot}&=&\frac{\rho}{2}\left\{\sum_{n=\ldots,-2,0,2,\ldots}\widetilde{u}\left(\frac{2\pi}{a}n\right)e^{-\frac{\pi^2}{\alpha}n^2}+\left(\frac{J(\alpha)}{I(\alpha)}\right)^2\sum_{n=\ldots,-3,-1,1,3,\ldots}\widetilde{u}\left(\frac{2\pi}{a}n\right)e^{-\frac{\pi^2}{\alpha}n^2}\right\}
\nonumber \\
&=&\frac{\rho}{2}\left\{\sum_{n=-\infty}^{\infty}\widetilde{u}\left(\frac{4\pi}{a}n\right)e^{-\frac{4\pi^2}{\alpha}n^2}+\left(\frac{J(\alpha)}{I(\alpha)}\right)^2\sum_{n=-\infty}^{\infty}\widetilde{u}\left[\frac{4\pi}{a}\left(n+\frac{1}{2}\right)\right]e^{-\frac{4\pi^2}{\alpha}\left(n+\frac{1}{2}\right)^2}\right\}\,,
\nonumber \\
\label{eq2-16}
\ea
with
\be
J(\alpha)=\sqrt{\frac{2\pi}{\alpha}}\sum_{n=-\infty}^{+\infty}e^{-\frac{2\pi^2}{\alpha}\left(n+\frac{1}{2}\right)^2}\,.
\label{eq2-17}
\ee
As $\alpha$ varies from 0 to infinity, the ratio $J(\alpha)/I(\alpha)$ grows monotonically from 0 to 1.

Instead of the ansatz (\ref{eq2-9}), we can directly express the square of $\psi$, that is $\eta(x)=V\psi^2(x)$ [see Eq.\,(\ref{eq2-7})], as a (normalized) sum of Gaussians centered on the lattice positions:
\be
{\rm VT}2:\,\,\,\,\,\,\eta(x)=\sqrt{\frac{2\alpha}{\pi}}\sum_{n=-\infty}^{+\infty}e^{-2\alpha\left(\frac{x}{a}-n\right)^2}
\label{eq2-18}
\ee
(notice the different form of the exponent with respect to (\ref{eq2-9}), made in order to provide a more meaningful comparison between VT1 and VT2 in the high-density limit). Since $\eta(x)$ is periodic, it can be expanded as a Fourier series:
\be
\eta(x)=\sum_G\eta_Ge^{iGx}\,.
\label{eq2-19}
\ee
Denoting ${\cal C}$ any crystalline cell and $R=na$, the Fourier coefficient $\eta_G$ is given by
\be
\eta_G=\frac{1}{a}\int_{\cal C}{\rm d}x\,e^{-iGx}\eta(x)=\frac{1}{V}\sqrt{\frac{2\alpha}{\pi}}\sum_R\underbrace{e^{-iGR}}_{1}\int_V{\rm d}x\,e^{-iG(x-R)}e^{-2\alpha\left(\frac{x-R}{a}\right)^2}=e^{-\frac{G^2a^2}{8\alpha}}\,.
\label{eq2-20}
\ee
Hence,
\be
\eta(x)=\sum_Ge^{-\frac{G^2a^2}{8\alpha}}e^{iGx}=\sum_Ge^{-\frac{G^2a^2}{8\alpha}}\cos(Gx)\,.
\label{eq2-21}
\ee
Normalization is clearly satisfied, since $(1/V)\int_V{\rm d}x\,\eta(x)=\sum_G\eta_G\delta_{G,0}=\eta_0=1$.

We now compute the specific potential energy:
\ba
{\cal E}_{\rm pot}&=&\frac{\rho}{2V}\int_V{\rm d}x\int_V{\rm d}x'\,\eta(x')u(x-x')\eta(x)
\nonumber \\
&=&\frac{\rho}{2V}\sum_{G,G'}\eta_G\eta_{G'}\underbrace{\int_V{\rm d}x'\,e^{i(G+G')x'}}_{V\delta_{G',-G}}\underbrace{\int_V{\rm d}x''\,e^{iGx''}u(x'')}_{\widetilde{u}(-G)=\widetilde{u}(G)}
\nonumber \\
&=&\frac{\rho}{2}\sum_G\widetilde{u}(G)\eta_G^2=\frac{\rho\widetilde{u}(0)}{2}+\rho\sum_{n=1}^\infty\widetilde{u}\left(\frac{2\pi}{a}n\right)e^{-\frac{\pi^2}{\alpha}n^2}\,.
\label{eq2-22}
\ea
For the PSM,
\be
\widetilde{u}(k)=2\epsilon\frac{\sin(k\sigma)}{k}\,\,\,\,\,\,{\rm and}\,\,\,\,\,\,{\cal E}_{\rm pot}=\rho\sigma\epsilon+\frac{\rho a}{\pi}\epsilon\sum_{n=1}^\infty\frac{1}{n}e^{-\frac{\pi^2}{\alpha}n^2}\sin\left(\frac{2\pi}{a}n\sigma\right)\,.
\label{eq2-23}
\ee
For any $u(x)$, ${\cal E}_{\rm pot}$ depends on $\alpha$ only through the quantity $e^{-\pi^2/\alpha}$, which increases monotonically from 0 to 1 when $\alpha$ runs from 0 to infinity.

%
%
\begin{figure}
\begin{center}
\includegraphics[width=15cm]{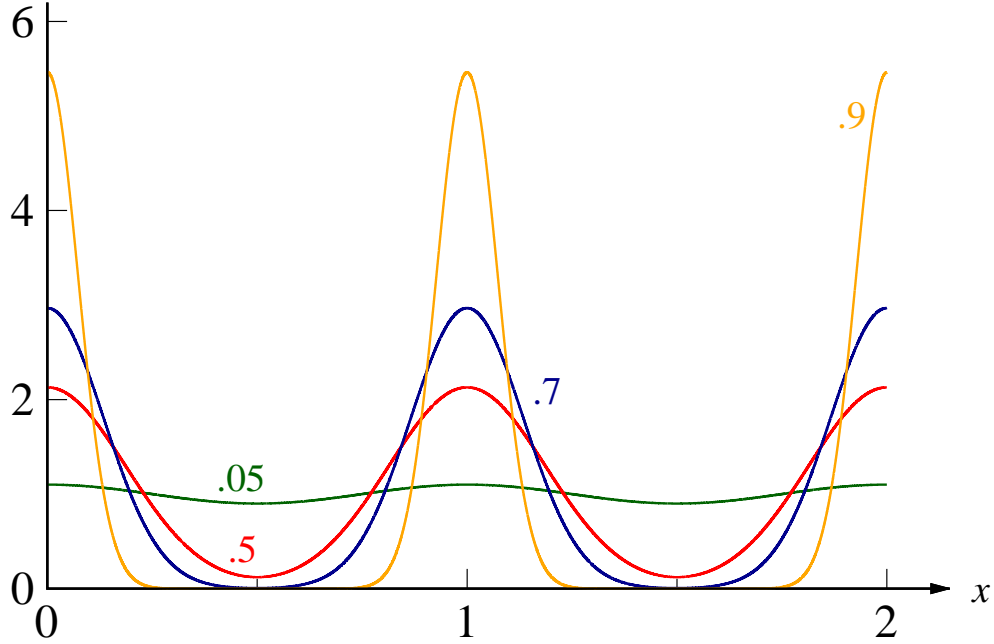}
\caption{$\vartheta_3(\pi x,q)$, for $0\le x\le 2$ and $q=0.05,0.5,0.7,0.9$.}
\label{fig1}
\end{center}
\end{figure}

The kinetic energy per particle is given by the first term in Eq.\,(\ref{eq2-8}):
\be
{\cal E}_{\rm kin}=\frac{\hbar^2}{8mV}\int_V{\rm d}x\,\frac{\eta^{\prime 2}(x)}{\eta(x)}=\frac{\hbar^2}{8ma}\int_{\cal C}{\rm d}x\,\frac{\eta^{\prime 2}(x)}{\eta(x)}\,.
\label{eq2-24}
\ee
To proceed further, one notices that $\eta(x)$ is intimately related to a Jacobi theta function [see Eqs.\,(\ref{eq2-21}) and (\ref{a-2})]:
\be
\eta(x)=\vartheta_3\left(\frac{\pi}{a}x,e^{-\frac{\pi^2}{2\alpha}}\right)\,.
\label{eq2-25}
\ee
A graph of this function is plotted in Fig.\,1. It is quite remarkable that a simple expression exists for the logarithmic derivative of $\vartheta_3$, see Eq.\,(\ref{a-4}). Using this formula we can write:
\be
\frac{\eta^{\prime 2}(x)}{\eta(x)}=\eta'(x)\frac{\eta'(x)}{\eta(x)}=-16\frac{\pi^2}{a^2}\sum_{n_1,n_2=1}^\infty(-1)^{n_1}n_2\frac{e^{-\frac{\pi^2}{2\alpha}(n_1+n_2^2)}}{1-e^{-\frac{\pi^2}{\alpha}n_1}}\sin\left(\frac{2\pi}{a}n_1x\right)\sin\left(\frac{2\pi}{a}n_2x\right)\,.
\label{eq2-26}
\ee
Since
\be
\int_{-a/2}^{a/2}{\rm d}x\,\sin\left(\frac{2\pi}{a}n_1x\right)\sin\left(\frac{2\pi}{a}n_2x\right)=\frac{a}{2}\delta_{n_1,n_2}\,,
\label{eq2-27}
\ee
we finally obtain:
\be
{\cal E}_{\rm kin}=\frac{\pi^2\sigma^2}{a^2}e_0\sum_{n=1}^\infty(-1)^{n-1}\frac{ne^{-\frac{\pi^2}{2\alpha}(n+n^2)}}{1-e^{-\frac{\pi^2}{\alpha}n}}\,.
\label{eq2-28}
\ee
Like ${\cal E}_{\rm pot}$, also ${\cal E}_{\rm kin}$ depends on $\alpha$ through $e^{-\pi^2/\alpha}$, while its $a$-dependence is simply ${\cal E}_{\rm kin}\propto a^{-2}$.

In the next Section we show that VT1 and VT2 give very similar energies; moreover, VT1 and VT2 also share with GP theory the same transition point.

\section{Results}
\setcounter{equation}{0}
\renewcommand{\thesubsection}{\arabic{subsection}}
\renewcommand{\theequation}{3.\arabic{equation}}

Here, we provide results obtained for a few instances of 1D softly-repulsive bosons at $T=0$ using three MF theories, namely GP theory and the variational theories introduced in Sec.\,II. All theories agree in predicting a continuous quantum transition from a fluid phase to a cluster-crystal phase. While the GP approximation provides by construction the best condensate wave function, i.e., the one with the lowest energy, we shall see that VT1 is indeed as accurate in describing the transition behavior as GP theory.

\subsection{Assessment of the variational approximations}
\setcounter{equation}{0}
\renewcommand{\thesubsection}{\arabic{subsection}}
\renewcommand{\theequation}{3.1.\arabic{equation}}

For any given value of $\rho$, we solve the system of equations (\ref{eq2-5}) for fixed $a$ by assuming that $c_K=0$ for $K=(2\pi/a)n$ and $|n|>8$ (nothing changes if this threshold were rather 12). We cyclically perform the diagonalization of the resulting $17\times 17$ Hermitian matrix of coefficients within the iterative procedure described in Sec.\,II, until self-consistency is reached. In the end, $a$ is optimized until its value is determined to five decimal places. Then, we solve VT1 and VT2, looking for the minimum of the energy functional on a grid of $(\alpha,a)$ values covering the region where the absolute minimum of ${\cal E}$ lies. The spacing of the grid is progressively reduced around the minimum, until its location is determined to $10^{-6}$ precision.

%
%
\begin{figure}
\begin{center}
\includegraphics[width=15cm]{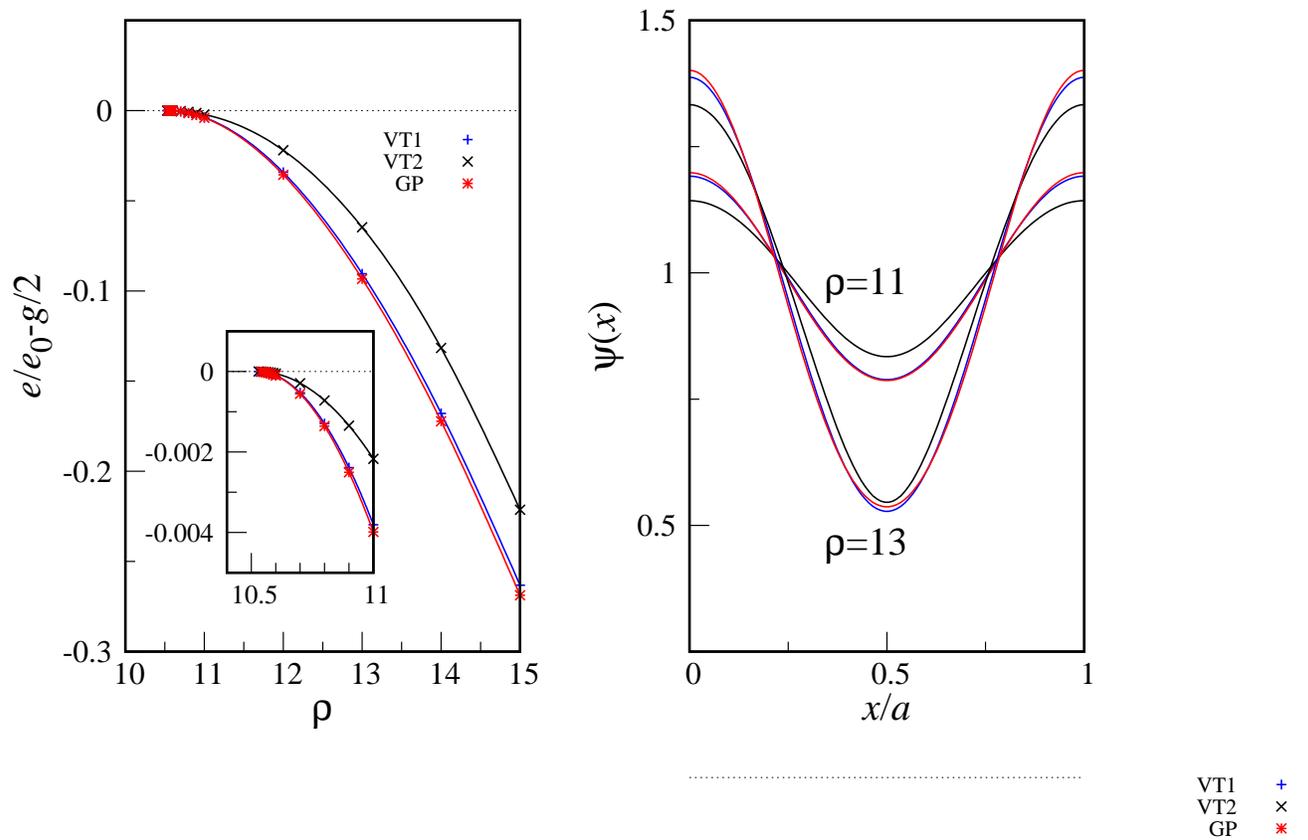}
\end{center}
\caption{1D PSM bosons at $T=0$. Left: excess energy (units of $e_0$) plotted as a function of the reduced density (in the inset, a magnification of the transition region is shown). Right: single-particle wave function for two reduced densities, $\rho=11$ and $\rho=13$ (red: GP theory; blue: VT1; black: VT2).}
\label{fig2}
\end{figure}

We show results for PSM bosons on a line in the left panel of Fig.\,2, where the excess energy $\Delta{\cal E}={\cal E}-\rho\sigma\epsilon$ is plotted as a function of $\rho$. Below $\rho_c\lesssim 10.53$ (units of $e_0\epsilon^{-1}\sigma^{-1}$), the minimum $\Delta{\cal E}$ is invariably zero for all theories (fluid phase); above $\rho_c$, the minimum excess energy is a negative number, and the system phase is a crystal (we better discuss the nature of this crystal in the following Sec.\,III.2). We observe that the shape of the best single-particle state is nearly identical for GP theory and VT1 --- see the right panel of Fig.\,2, where the condensate wave functions for all theories are plotted side by side for $\rho=11$ and 13.

We have verified that the same degree of similarity between GP theory and VT1 also holds for the softened van der Waals (SVDW) repulsion, $u(r)=\epsilon/[1+(r/\sigma)^6]$, which is the same interaction investigated in Ref.\,\cite{Rossotti}. Again, the transition threshold turns out to be the same in both theories ($\rho_c\lesssim 20.65$).

\subsection{Analysis of the transition region}
\setcounter{equation}{0}
\renewcommand{\theequation}{3.2.\arabic{equation}}

The results of Sec.\,III.1 indicate that the phase-transition threshold at $T=0$ of 1D penetrable bosons is by all evidence identical in GP theory and VT1 (we provide an explanation of this fact at the end of this Section). To better inquire into the system behavior near the transition, as well as to uncover similarities and differences between the various approaches, we derive below a small-$\alpha$ expansion of the VT1 and VT2 energy functionals that suffices for all purposes.

Let us first consider VT1. To calculate the energy [Eqs.\,(\ref{eq2-15}) and (\ref{eq2-16})], we need the perturbative expansions of $I(\alpha)$ [Eq.\,(\ref{eq2-14})] and $J(\alpha)$ [Eq.\,(\ref{eq2-17})] around $\alpha=0$. In this respect, it comes useful to express these functions in terms of Jacobi theta functions:
\ba
I(\alpha)=\sqrt{\frac{2\pi}{\alpha}}\sum_{n=-\infty}^{+\infty}e^{-\frac{2\pi^2}{\alpha}n^2}=\sqrt{\frac{2\pi}{\alpha}}\vartheta_3(0,e^{-\frac{2\pi^2}{\alpha}})\,;
\nonumber \\
J(\alpha)=\sqrt{\frac{2\pi}{\alpha}}\sum_{n=-\infty}^{+\infty}e^{-\frac{2\pi^2}{\alpha}\left(n+\frac{1}{2}\right)^2}=\sqrt{\frac{2\pi}{\alpha}}\vartheta_2(0,e^{-\frac{2\pi^2}{\alpha}})\,.
\label{eq3-2-1}
\ea
Combining Eqs.\,(\ref{eq3-2-1}) and (\ref{a-3}) we obtain:
\ba
\left(\frac{J(\alpha)}{I(\alpha)}\right)^2&=&4e^{-\frac{\pi^2}{\alpha}}\left(1-4e^{-\frac{2\pi^2}{\alpha}}+14e^{-\frac{4\pi^2}{\alpha}}+\ldots\right)\,;
\nonumber \\
\frac{I'(\alpha)}{I(\alpha)}&=&-\frac{1}{2\alpha}+\frac{4\pi^2}{\alpha^2}e^{-\frac{2\pi^2}{\alpha}}\left(1-2e^{-\frac{2\pi^2}{\alpha}}+4e^{-\frac{4\pi^2}{\alpha}}+\ldots\right)\,,
\label{eq3-2-2}
\ea
whence the following small-$\alpha$ expansion of the difference in energy between the crystalline and fluid solutions:
\ba
\Delta{\cal E}&\equiv&{\cal E}_{\rm kin}+{\cal E}_{\rm pot}-\rho\sigma\epsilon=4\left[\frac{\pi^2\sigma^2}{a^2}e_0+\rho\widetilde{u}\left(\frac{2\pi}{a}\right)\right]e^{-\frac{2\pi^2}{\alpha}}
\nonumber \\
&+&\left\{-\frac{8\pi^2\sigma^2}{a^2}e_0+\rho\left[\widetilde{u}\left(\frac{4\pi}{a}\right)-16\widetilde{u}\left(\frac{2\pi}{a}\right)\right]\right\}e^{-\frac{4\pi^2}{\alpha}}
\nonumber \\
&+&8\left[\frac{2\pi^2\sigma^2}{a^2}e_0+7\rho\widetilde{u}\left(\frac{2\pi}{a}\right)\right]e^{-\frac{6\pi^2}{\alpha}}+\ldots
\label{eq3-2-3}
\ea

For the sake of clarity, let us consider the case of PSM bosons. For small $\alpha$, an approximation sufficient for the analysis of the transition behavior is:
\be
\Delta{\cal E}(X,a;\rho)\simeq rX^2+wX^4\,,
\label{eq3-2-4}
\ee
with $X=e^{-\pi^2/\alpha}$ and
\ba
r&=&4\left[\frac{\pi^2\sigma^2}{a^2}e_0+\frac{\rho a}{\pi}\epsilon\sin\left(\frac{2\pi}{a}\sigma\right)\right]\,;
\nonumber \\
w&=&-\frac{8\pi^2\sigma^2}{a^2}e_0+\frac{\rho a}{\pi}\epsilon\left[\frac{1}{2}\sin\left(\frac{4\pi}{a}\sigma\right)-16\sin\left(\frac{2\pi}{a}\sigma\right)\right]\,.
\label{eq3-2-5}
\ea
Notice that $r$ and $w$ are explicit functions of $a$ and $\rho$. The extremal points of $\Delta{\cal E}(X)$ are $\overline{X}=0$ and (if $r<0$) the non-zero root of $\Delta{\cal E}'(X)=0$, that is $\overline{X}=\sqrt{-r/(2w)}$, with specific energies $\Delta{\cal E}_{\rm min}(a,\rho)=0$ and $-r^2/(4w)$, respectively (it turns out that $w>0$ in the relevant range of $a$ and $\rho$ values). The non-trivial solution exists providing that $r<0$, namely $\rho>\rho_0$, with
\be
\rho_0\sigma=-\frac{e_0}{\epsilon}\frac{\pi^3}{\left(\frac{a}{\sigma}\right)^3\sin\left(\frac{2\pi}{a}\sigma\right)}\,.
\label{eq3-2-6}
\ee
Since $r=4\widetilde{u}(2\pi/a)(\rho-\rho_0)$, it follows that
\be
\Delta{\cal E}_{\rm min}=-\frac{r^2}{4w}=-\frac{4\widetilde{u}^2(2\pi/a)}{w}(\rho-\rho_0)^2\,.
\label{eq3-2-7}
\ee
The equilibrium lattice constant, $\overline{a}(\rho)$, is the one providing the minimum value of $\Delta{\cal E}_{\rm min}(a)$ for the given $\rho$.

In order that $\rho_0>0$, it is sufficient that $1<a/\sigma<2$; in this interval, $w>0$ as well (in fact, there are infinite other intervals where $\sin(2\pi\sigma/a)<0$, namely $1/2<a/\sigma<2/3,1/3<a/\sigma<2/5,\ldots$, but these other $a$ provide much larger values of $\rho_0$ [see Eq.\,(\ref{eq3-2-6})] and, above this density, also energy minima higher than those in the interval $1<a/\sigma<2$). The smallest density above which the energy is negative is the minimum of $\rho_0(a)$, i.e., $\rho_0(a_c)\equiv\rho_c$. For $y=-x^3\sin(2\pi/x)$, the derivative $y'\ge 0$ for
\be
\frac{x}{2\pi}\tan\frac{2\pi}{x}\ge\frac{1}{3}\Longrightarrow x\le 1.540695087\ldots\equiv\frac{a_c}{\sigma}
\label{eq3-2-8}
\ee
(the other roots of Eq.\,(\ref{eq3-2-8}) fall outside the range from 1 to 2). Hence, the transition from fluid to crystal occurs for $\rho=\rho_c=10.524990629\ldots$; at this density, $\overline{X}$ switches continuously from 0 (stable fluid) to $\sqrt{-r/(2w)}\propto\sqrt{(\rho-\rho_c)/\rho_c}$ (stable crystal). For densities larger than $\rho_c$, the minimum of energy occurs at a $\overline{a}$ smaller than $a_c$, see Fig.\,3. Slightly above $\rho_c$, the behavior of $\overline{\alpha}$ and of the excess energy are, up to a ${\cal O}(1)$ factor:
\be
\overline{\alpha}(\rho)\sim\left|\ln\left(\frac{\rho-\rho_c}{\rho_c}\right)\right|^{-1}\,\,\,\,\,\,{\rm and}\,\,\,\,\,\,\Delta e(\rho)\sim -e_0\left(\frac{\rho-\rho_c}{\rho_c}\right)^2\,.
\label{eq3-2-9}
\ee
In particular, the order parameter $\overline{\alpha}$ is continuous for $\rho=\rho_c$. At the transition, the crystal pressure equals that of the fluid, as it follows from the general relation $P(\rho)=\rho^2 e'(\rho)$ and the second of Eqs.\,(\ref{eq3-2-9}). The transition pressure is $P_c=\sigma\epsilon\rho_c^2=110.7754\ldots$ (units of $e_0^2\epsilon^{-1}\sigma^{-1}$). The phase transition is continuous since the energy and its derivative are continuous at $\rho=\rho_c$ (the phase with the minimum enthalpy is the one with the minimum energy). At the transition, the average number of particles per site is $\rho_ca_c=16.2157\ldots$ (hence, the dense phase is a cluster crystal).

%
%
\begin{figure}
\begin{center}
\includegraphics[width=15cm]{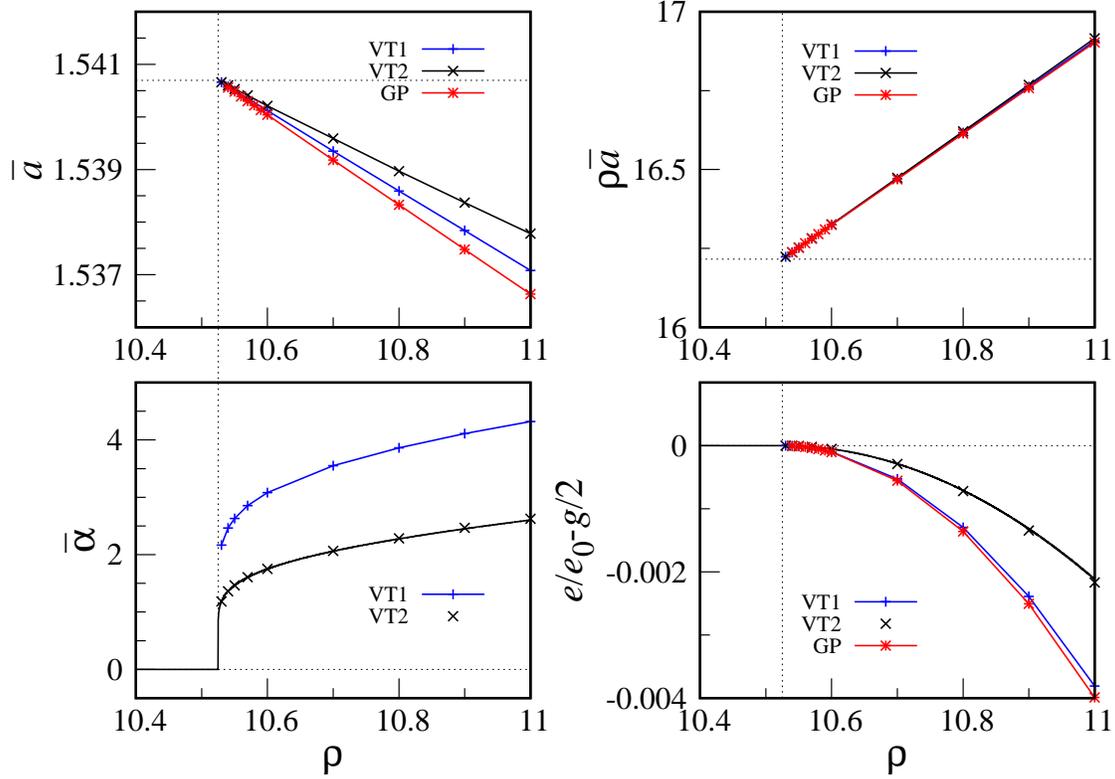}
\end{center}
\caption{Transition behavior of 1D PSM bosons at $T=0$: comparison between GP theory (red), VT1 (blue), and VT2 (black). We report numerical results (symbols joined by straight-line segments) and, only for VT2, also theoretical results (the black lines in the bottom panels, see Eq.\,(\ref{eq3-2-11})). Top left: lattice constant. Top right: the quantity $\rho a$, representing the average number of particles in a cluster. Bottom left: best value of $\alpha$. Bottom right: excess energy. The dotted lines mark the transition values.}
\label{fig3}
\end{figure}

Similar results hold for VT2. It follows from Eqs.\,(\ref{eq2-23}) and (\ref{eq2-28}) that
\ba
\Delta{\cal E}&\equiv&{\cal E}_{\rm kin}+{\cal E}_{\rm pot}-\rho\sigma\epsilon
\nonumber \\
&=&\left[\frac{\pi^2\sigma^2}{a^2}e_0+\frac{\rho a}{\pi}\epsilon\sin\left(\frac{2\pi}{a}\sigma\right)\right]e^{-\frac{\pi^2}{\alpha}}+\frac{\pi^2\sigma^2}{a^2}e_0\left(e^{-\frac{2\pi^2}{\alpha}}-e^{-\frac{3\pi^2}{\alpha}}\right)
\nonumber \\
&+&\left[\frac{\pi^2\sigma^2}{a^2}e_0+\frac{\rho a}{2\pi}\epsilon\sin\left(\frac{4\pi}{a}\sigma\right)\right]e^{-\frac{4\pi^2}{\alpha}}+\ldots
\label{eq3-2-10}
\ea
As before, for small values of $X=\exp\{-\pi^2/(2\alpha)\}$ (which is different from the definition of $X$ given before) the same approximation (\ref{eq3-2-4}) holds, with $w>0$ independent of the density and a $r$ value which is one fourth of that for VT1. This implies that $\rho_c$ and $a_c$ are identical for VT1 and VT2. In particular, near $\rho=\rho_c$ the behavior of $\overline{\alpha}$ and of the excess energy are the following:
\be
\overline{\alpha}(\rho)\sim\frac{\pi^2}{\left|\ln\left(\frac{\rho-\rho_c}{2\rho_c}\right)\right|}\,\,\,\,\,\,{\rm and}\,\,\,\,\,\,\Delta e(\rho)\sim -\frac{\pi^2\sigma^2}{4\rho_c^2a_c^2}e_0(\rho-\rho_c)^2\,.
\label{eq3-2-11}
\ee
We find an unexpected outcome when computing the isothermal compressibility at $T=0$,
\ba
K_T^{-1}&=&-\left.V\frac{\partial P}{\partial V}\right|_{T=0}=\rho P'(\rho)=2\rho^2 e'(\rho)+\rho^3 e''(\rho)\,.
\label{eq3-2-12}
\ea
For $\rho=\rho_c$, $K_T^{-1}$ has the following values in the two phases:
\be
{\rm F}:\,\,\,K_T^{-1}=2(\rho_c\sigma)^2\frac{\epsilon}{\sigma}\,;\,\,\,\,\,\,\,\,\,\,{\rm C}:\,\,\,K_T^{-1}=2(\rho_c\sigma)^2\frac{\epsilon}{\sigma}-\frac{\pi^2}{2}\left(\frac{\sigma}{a_c}\right)^2(\rho_c\sigma)\frac{e_0}{\sigma}\,.
\label{eq3-2-13}
\ee
Hence, $K_T$ undergoes a jump at the transition and, right at the transition point, the crystal is more compressible than the fluid.

Summarizing, all three theories predict the same transition density $\rho_c$, whereas small differences arise for $\rho>\rho_c$ (see Fig.\,3). We can appreciate from this figure that the VT1 approximation is superior to VT2, since it gives a smaller energy; moreover, VT1 has practically the same accuracy of GP theory.

To explain why VT1 and VT2 give exactly the same transition density as GP theory, we consider the nature of fluid excitations in the latter theory. As well known (see, e.g., \cite{Kunimi,Prestipino3,Zloshchastiev}), the spectrum of these excitations is Bogoliubov-like:
\be
\hbar\omega(k)=\sqrt{\frac{\hbar^2 k^2}{2m}\left(\frac{\hbar^2 k^2}{2m}+2\rho\widetilde{u}(k)\right)}\,.
\label{eq3-2-14}
\ee
According to this dispersion law, if $\widetilde{u}(k)$ is negative in some range of $k$ (as occurs for any bounded interaction that is ``fatter'' than Gaussian) then the system is superfluid by Landau's argument. In this case, above a certain density a maxon peak develops in $\omega(k)$, followed at larger $k$ by a roton minimum; increasing the density further, the roton eventually softens and the fluid becomes unstable towards the formation of a density wave (this happens when the quantity within parentheses in Eq.\,(\ref{eq3-2-14}) vanishes). Putting $k=2\pi/a$, roton softening first occurs for a density and a value of the wavelength $a$ such that
\be
\frac{\pi^2\sigma^2}{a^2}e_0+\rho\widetilde{u}\left(\frac{2\pi}{a}\right)=0\,\,\,{\rm and}\,\,\,\widetilde{u}\left(\frac{2\pi}{a}\right)=\frac{\pi}{a}\widetilde{u}^\prime\left(\frac{2\pi}{a}\right)\,,
\label{eq3-2-15}
\ee
where the rationale behind the second equation is that the density $\rho_0(a)$ solving the first equation be as low as possible. These are exactly the same conditions for the occurrence of clusterization in variational theory. In particular, for the SVDW potential the Fourier transform reads
\be
\widetilde{u}(k)=\frac{\pi}{3}\epsilon\,e^{-k/2}\left[e^{-k/2}+\cos\left(\frac{\sqrt{3}}{2}k\right)+\sqrt{3}\sin\left(\frac{\sqrt{3}}{2}k\right)\right]
\label{eq3-2-16}
\ee
(this is obtained by evaluating with the residue theorem the integral of $e^{ikz}/(1+z^6)$ over a large semi-circular contour inscribed in the ${\rm Re}\,z\ge 0$ half-plane). By numerically solving Eqs.\,(\ref{eq3-2-15}) we confirm that clusterization of SVDW bosons in MF theory occurs at $\rho_c=20.64654\ldots$, which is the same condition for roton softening in the $\epsilon\rightarrow 0$ limit quoted by Rossotti and coworkers. It is worth observing that the MF locus for roton softening gives a good approximation to the exact transition line even far away from the MF limit (see Fig.\,1 in Ref.\,\cite{Rossotti}).

\subsection{PSM bosons: High-density limit}
\setcounter{equation}{0}
\renewcommand{\theequation}{3.3.\arabic{equation}}

After highlighting the features of clusterization at $T=0$ of 1D soft-core bosons through an analysis of the small-$\alpha$ limit of the energy functional, we now consider the opposite limit of very large $\alpha$ values, which corresponds to high densities, focusing on the case of PSM bosons. It turns out that a duality property of Jacobi theta functions makes this limit accessible analytically.

Starting with VT1, we first approximate the kinetic energy (\ref{eq2-15}) by noting that, for $\alpha\gg 1$,
\be
I(\alpha)=\sum_{n=-\infty}^{+\infty}e^{-\frac{\alpha}{2}n^2}\sim 1+2e^{-\frac{\alpha}{2}}\,.
\label{eq3-3-1}
\ee
We immediately obtain:
\be
{\cal E}_{\rm kin}\sim \frac{\alpha\sigma^2}{2a^2}e_0\left(1-2\alpha e^{-\frac{\alpha}{2}}\right)\,.
\label{eq3-3-2}
\ee
Then, we estimate the ratio $[J(\alpha)/I(\alpha)]^2$ appearing in the potential-energy formula, Eq.\,(\ref{eq2-16}). From the second of Eqs.\,(\ref{eq3-2-1}), using a duality property of theta functions [Eq.\,(\ref{a-5})] we obtain:
\be
J(\alpha)=\vartheta_4\left(0,e^{-\frac{\alpha}{2}}\right)=\sum_{n=-\infty}^{+\infty}(-1)^ne^{-\frac{\alpha}{2}n^2}\sim 1-2e^{-\frac{\alpha}{2}}\,,
\label{eq3-3-3}
\ee
so that
\be
\left(\frac{J(\alpha)}{I(\alpha)}\right)^2\sim1-8e^{-\frac{\alpha}{2}}\,.
\label{eq3-3-4}
\ee

Next, we evaluate for PSM bosons the sum
{\small
\be
\sum_{n=-\infty}^{+\infty}\widetilde{u}\left(\frac{4\pi}{a}n\right)e^{-\frac{4\pi^2}{\alpha}n^2}=2\sigma\epsilon+\frac{2\sigma\epsilon}{z}\sum_{n=1}^{\infty}\frac{1}{n}e^{-\frac{4\pi^2}{\alpha}n^2}\sin(2nz)\equiv 2\sigma\epsilon+\frac{2\sigma\epsilon}{z}f(z)\,,
\label{eq3-3-5}
\ee
}
with $z=2\pi\sigma/a$. In order to estimate the large-$\alpha$ limit of Eq.\,(\ref{eq3-3-5}), we note that:
\be
f'(z)=\vartheta_3\left(z,e^{-\frac{4\pi^2}{\alpha}}\right)-1\,.
\label{eq3-3-6}
\ee
By another duality formula [Eq.\,(\ref{a-6})] we get:
\ba
\vartheta_3\left(z,e^{-\frac{4\pi^2}{\alpha}}\right)&=&\sqrt{\frac{\alpha}{4\pi}}e^{-\frac{\alpha}{4\pi^2}z^2}\vartheta_3\left(i\frac{\alpha z}{4\pi},e^{-\frac{\alpha}{4}}\right)
\nonumber \\
&=&\sqrt{\frac{\alpha}{4\pi}}e^{-\frac{\alpha}{4\pi^2}z^2}\left[1+2\sum_{n=1}^\infty e^{-\frac{\alpha}{4}n^2}\cosh\left(\frac{\alpha z}{2\pi}n\right)\right]\,.
\label{eq3-3-7}
\ea
For $1\le a/\sigma\le 2$, the value of $z$ is between $\pi$ and $2\pi$. In this range, the leading terms in the expansion (\ref{eq3-3-7}) are (in equal measure) the first and the second one (while the 1 can be ignored), thus obtaining:
\be
f'(z)\sim\sqrt{\frac{\alpha}{4\pi}}\left[e^{-\frac{\alpha}{4}\left(\frac{z}{\pi}-1\right)^2}+e^{-\frac{\alpha}{4}\left(2-\frac{z}{\pi}\right)^2}\right]-1\,.
\label{eq3-3-8}
\ee
Noting that $f(\pi)=0$, we have:
\be
f(z)\sim\frac{\alpha}{4\pi}\int_\pi^z{\rm d}t\left[e^{-\frac{\alpha}{4}\left(\frac{t}{\pi}-1\right)^2}+e^{-\frac{\alpha}{4}\left(\frac{t}{\pi}-2\right)^2}\right]-(z-\pi)\,.
\label{eq3-3-9}
\ee
The integral returns error functions, whose limiting behavior for large values of the argument is:
\be
{\rm erf}(x)\equiv\frac{2}{\sqrt{\pi}}\int_0^x{\rm d}t\,e^{-t^2}\sim 1-\frac{e^{-x^2}}{\sqrt{\pi}x}\,.
\label{eq3-3-10}
\ee
After obvious steps we eventually find:
{\small
\be
\sum_{n=-\infty}^{+\infty}\widetilde{u}\left(\frac{4\pi}{a}n\right)e^{-\frac{4\pi^2}{\alpha}n^2}\sim a\epsilon+\frac{a\epsilon}{\pi}\left\{\frac{\pi}{2}-\sqrt{\frac{\pi}{\alpha}}\left[e^{-\frac{\alpha}{4}}+\frac{e^{-\frac{\alpha}{4}\left(\frac{2\sigma}{a}-1\right)^2}}{\frac{2\sigma}{a}-1}+\frac{e^{-\frac{\alpha}{4}\left(\frac{2\sigma}{a}-2\right)^2}}{\frac{2\sigma}{a}-2}\right]\right\}\,.
\label{eq3-3-11}
\ee
}

Then, we examine the asymptotic behavior of
\ba
&&\sum_{n=-\infty}^{+\infty}\widetilde{u}\left[\frac{4\pi}{a}\left(n+\frac{1}{2}\right)\right]e^{-\frac{4\pi^2}{\alpha}\left(n+\frac{1}{2}\right)^2}
\nonumber \\
&=&\frac{2\sigma\epsilon}{z}\sum_{n=-\infty}^{\infty}\frac{1}{2n+1}e^{-\frac{4\pi^2}{\alpha}\left(n+\frac{1}{2}\right)^2}\sin\left[(2n+1)z\right]\equiv\frac{2\sigma\epsilon}{z}f(z)\,,
\label{eq3-3-12}
\ea
with $z=2\pi\sigma/a$. By the same above considerations, we obtain:
\ba
f'(z)&=&\vartheta_2\left(z,e^{-\frac{4\pi^2}{\alpha}}\right)=\sqrt{\frac{\alpha}{4\pi}}e^{-\frac{\alpha}{4\pi^2}z^2}\left[1+2\sum_{n=1}^\infty(-1)^ne^{-\frac{\alpha}{4}n^2}\cosh\left(\frac{\alpha z}{2\pi}n\right)\right]
\nonumber \\
&\sim&\sqrt{\frac{\alpha}{4\pi}}\left[-e^{-\frac{\alpha}{4}\left(\frac{z}{\pi}-1\right)^2}+e^{-\frac{\alpha}{4}\left(2-\frac{z}{\pi}\right)^2}\right]\,.
\label{eq3-3-13}
\ea
Integrating (\ref{eq3-3-13}) from $\pi$ and $z$, and then plugging the result in (\ref{eq3-3-12}), we arrive at:
\ba
&&\sum_{n=-\infty}^{+\infty}\widetilde{u}\left[\frac{4\pi}{a}\left(n+\frac{1}{2}\right)\right]e^{-\frac{4\pi^2}{\alpha}\left(n+\frac{1}{2}\right)^2}
\nonumber \\
&&\sim\frac{a\epsilon}{\pi}\left\{-\frac{\pi}{2}-\sqrt{\frac{\pi}{\alpha}}\left[e^{-\frac{\alpha}{4}}-\frac{e^{-\frac{\alpha}{4}\left(\frac{2\sigma}{a}-1\right)^2}}{\frac{2\sigma}{a}-1}+\frac{e^{-\frac{\alpha}{4}\left(\frac{2\sigma}{a}-2\right)^2}}{\frac{2\sigma}{a}-2}\right]\right\}\,.
\label{eq3-3-14}
\ea
Putting Eqs.\,(\ref{eq3-3-2}), (\ref{eq3-3-4}), (\ref{eq3-3-11}), and (\ref{eq3-3-14}) together, we obtain the sought-for high-density approximation of the VT1 energy functional.

The treatment is simpler for VT2. As far as kinetic energy is concerned, we start from $\eta(x)=\vartheta_3[\pi x/a,\exp\{-\pi^2/(2\alpha)\}]$. Observing that
\be
\vartheta_3\left(z,e^{-\frac{\pi^2}{2\alpha}}\right)=\sqrt{\frac{2\alpha}{\pi}}e^{-\frac{2\alpha}{\pi^2}z^2}\left[1+2\sum_{n=1}^\infty e^{-2\alpha n^2}\cosh\left(\frac{4\alpha z}{\pi}n\right)\right]\,,
\label{eq3-3-15}
\ee
for $-a/2\le x\le a/2$ we obtain
\be
\eta(x)\sim\sqrt{2\alpha}{\pi}e^{-\frac{2\alpha}{a^2}x^2}\,, 
\label{eq3-3-16}
\ee
whence we find [cf. Eq.\,(\ref{eq2-24})]
\be
{\cal E}_{\rm kin}=\frac{\alpha\sigma^2}{2a^2}e_0\,.
\label{eq3-3-17}
\ee

As for the potential energy, for PSM bosons it equals [see Eq.\,(\ref{eq2-23})]
\be
{\cal E}_{\rm pot}=\rho\sigma\epsilon+\frac{\rho\sigma\epsilon}{z}\sum_{n=1}^\infty\frac{1}{n}e^{-\frac{\pi^2}{\alpha}n^2}\sin(2nz)\equiv\frac{\rho\sigma\epsilon}{z}f(z)\,,
\label{eq3-3-18}
\ee
with $z=\pi\sigma/a$ and
\be
f'(z)=\vartheta_3\left(z,e^{-\frac{\pi^2}{\alpha}}\right)-1\,.
\label{eq3-3-19}
\ee
On the other hand,
\be
\vartheta_3\left(z,e^{-\frac{\pi^2}{\alpha}}\right)=\sqrt{\frac{\alpha}{\pi}}e^{-\frac{\alpha}{\pi^2}z^2}\left[1+2\sum_{n=1}^\infty e^{-\alpha n^2}\cosh\left(\frac{2\alpha z}{\pi}n\right)\right]\,.
\label{eq3-3-20}
\ee
For $1\le a/\sigma\le2$, $z$ falls between $\pi/2$ and $\pi$. In this interval, the leading term for $\alpha\gg 1$ is the first one in the expansion (\ref{eq3-3-20}), while the 1 can be ignored, thus arriving at:
\be
f'(z)\sim\sqrt{\frac{\alpha}{\pi}}e^{-\alpha\left(1-\frac{z}{\pi}\right)^2}-1\,.
\label{eq3-3-21}
\ee
Integrating from $\pi/2$ (where $f$ vanishes) and $z$, we thus obtain:
\be
f(z)\sim\sqrt{\frac{\pi}{4\alpha}}\left[\frac{e^{-\alpha\left(1-\frac{z}{\pi}\right)^2}}{1-\frac{z}{\pi}}-2e^{-\frac{\alpha}{4}}\right]-\left(z-\frac{\pi}{2}\right)
\label{eq3-3-22}
\ee
and
\be
{\cal E}_{\rm pot}\sim\frac{\rho a\epsilon}{2}+\frac{\rho a\epsilon}{\pi}\sqrt{\frac{\pi}{4\alpha}}\left[\frac{e^{-\alpha\left(1-\frac{\sigma}{a}\right)^2}}{1-\frac{\sigma}{a}}-2e^{-\frac{\alpha}{4}}\right]\,.
\label{eq3-3-23}
\ee
The sum of Eqs.\,(\ref{eq3-3-17}) and (\ref{eq3-3-23}) is the asymptotic expression of the VT2 energy functional.

%
%
\begin{figure}
\begin{center}
\includegraphics[width=15cm]{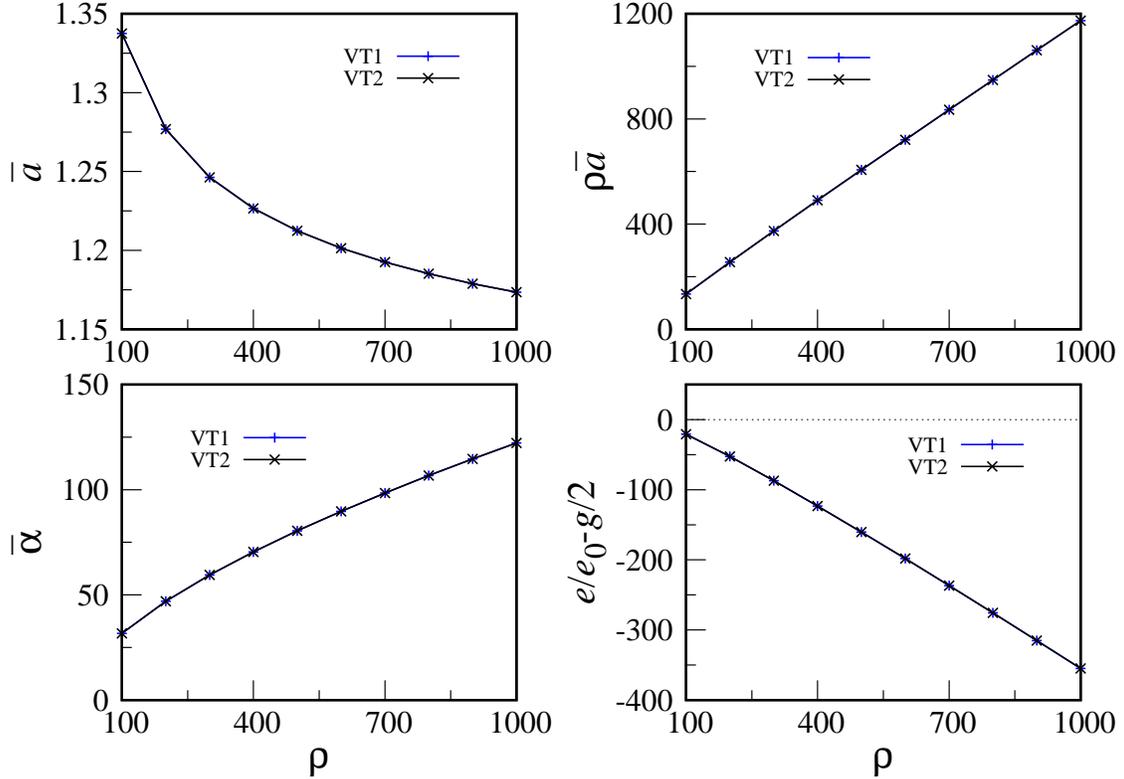}
\end{center}
\caption{High-density behavior of 1D PSM bosons at $T=0$: VT1 (blue) and VT2 data (black), obtained using the simplified energy functionals of Sec.\,III.4. Top left: lattice constant. Top right: the quantity $\rho a$, representing the average number of particles in a cluster. Bottom left: best value of $\alpha$. Bottom right: excess energy.}
\label{fig4}
\end{figure}

Using the simplified energy functionals, we obtain the data plotted in Fig.\,4. It turns out that, for high densities, VT1 and VT2 give practically the same optimal values of the variational parameters. For all densities above $\approx 80$, these values are hardly distinguishable from those extracted from the original functionals. In particular, VT1 energy is imperceptibly smaller than VT2 energy.

\subsection{PSM bosons: Supersolidity of the cluster phase}
\setcounter{equation}{0}
\renewcommand{\theequation}{3.4.\arabic{equation}}

Consider a quantum solid of cylindrical shape, set in uniform rotation around the axis. Leggett~\cite{Leggett1,Leggett2} has proposed to call {\em superfluid fraction} of the system the quantity
\be
f_s=\frac{I_0-I}{I_0}\,,
\label{eq3-4-1}
\ee
where $I$ is the moment of inertia around the cylinder axis and $I_0$ its classical value. It turns out that, like a superfluid, also a quantum solid may exhibit an anomalous response to axial rotations: for low rotation speed, part of the solid may stand still, with the result that $I<I_0$ and therefore $f_s>0$. In this case, the system is called a {\em supersolid}. The specificity of 1D is that, strictly speaking, an infinite crystal does not exist at $T=0$; however, the notion of supersolidity may still be considered for a finite, but large, 1D system in the CLL phase, e.g., for clusters arranged in a ring that rotates around its axis. 

Leggett has derived an exact formula for the superfluid fraction of a 1D solid of identical bosons at $T=0$~\cite{Leggett2}:
\be
f_s=\left(\frac{1}{a}\int_{\cal C}{\rm d}x\frac{1}{\eta(x)}\right)^{-1}\,,
\label{eq3-4-2}
\ee
where $a$ is the volume of the crystalline cell ${\cal C}$ and $\eta(x)=V\psi^2(x)$. The same result (\ref{eq3-4-2}) has been obtained by Sepulveda {\em et al.} within an approximate theory of the supersolid phase~\cite{Sepulveda}.

For the sake of clarity, let us take PSM bosons in 1D. We first consider VT2, for which $\eta(x)=\vartheta_3(\pi x/a,\exp\{-\pi^2/(2\alpha)\})$. At low density, we can write:
{\footnotesize
\ba
\frac{1}{\vartheta_3\left(\pi x/a,e^{-\frac{\pi^2}{2\alpha}}\right)}&=&1-2e^{-\frac{\pi^2}{2\alpha}}\cos\left(\frac{2\pi}{a}x\right)+4e^{-\frac{\pi^2}{\alpha}}\cos^2\left(\frac{2\pi}{a}x\right)-8e^{-\frac{3\pi^2}{2\alpha}}\cos^3\left(\frac{2\pi}{a}x\right)
\nonumber \\
&+&2e^{-\frac{2\pi^2}{\alpha}}\left[8\cos^4\left(\frac{2\pi}{a}x\right)-\cos\left(\frac{4\pi}{a}x\right)\right]+{\cal O}\left(e^{-\frac{5\pi^2}{2\alpha}}\right)\,.
\label{eq3-4-3}
\ea
}
Hence, we find:
\be
\frac{1}{a}\int_0^a{\rm d}x\frac{1}{\eta(x)}=1+2e^{-\frac{\pi^2}{\alpha}}+6e^{-\frac{2\pi^2}{\alpha}}+{\cal O}\left(e^{-\frac{5\pi^2}{2\alpha}}\right)\,,
\label{eq3-4-4}
\ee
and finally:
\be
f_s=1-2e^{-\frac{\pi^2}{\alpha}}-2e^{-\frac{2\pi^2}{\alpha}}+\ldots
\label{eq3-4-5}
\ee
This indicates that, close to the transition point, the superfluid fraction of the crystal varies as [cf. the first of Eqs.\,(\ref{eq3-2-11})]:
\be
f_s=1-\frac{\rho-\rho_c}{\rho_c}-\frac{(\rho-\rho_c)^2}{2\rho_c^2}+\ldots
\label{eq3-4-6}
\ee
Hence, the crystalline solid is a {\em supersolid} whose superfluid fraction is exactly one at the transition, then reducing progressively on compression. It is natural to ask whether $f_s$ eventually vanishes at a certain large value of the density, and the system then becomes a normal solid. We shall see that the answer is in the negative, at least within variational theory.

In the limit of high densities, Eq.\,(\ref{eq3-3-15}) allows us to write $1/\eta(x)$ as the ratio between two large quantities. However, we were not able to put this ratio in the form of a rapidly-convergent series. It is much simpler to obtain a positive lower limit for $f_s$, and this way conclude that the crystalline solid is, like in higher dimensions~\cite{Prestipino3}, a supersolid also for very large density. Called $\eta_{\rm min}=\eta(a/2)$ and $\eta_{\rm max}=\eta(0)$ the minimum and maximum value of $\eta(x)$ in the cell, we can write:
\be
f_s=\left[\frac{1}{a}\int_0^a{\rm d}x\,\eta(x)\cdot\frac{1}{a}\int_0^a{\rm d}x\frac{1}{\eta(x)}\right]^{-1}\ge\frac{\eta_{\rm min}}{\eta_{\rm max}}\,.
\label{eq3-4-7}
\ee
We derive from Eq.\,(\ref{eq3-3-16}) that:
\be
\eta_{\rm min}\sim 2\sqrt{2\overline{\alpha}}\pi e^{-\frac{\overline{\alpha}}{2}}\,\,\,\,\,\,{\rm and}\,\,\,\,\,\,\eta_{\rm max}\sim \sqrt{2\overline{\alpha}}\pi\,,
\label{eq3-4-8}
\ee
whence it follows:
\be
f_s\ge 2e^{-\frac{\overline{\alpha}}{2}}>0\,,
\label{eq3-4-9}
\ee
as anticipated.

In VT1, where $\sqrt{V}\psi(x)=C_\alpha'\vartheta_3[\pi x/a,\exp(-\pi^2/\alpha)]$ and $\eta(x)=V\psi^2(x)$, near the transition point the following expansion holds:
{\footnotesize
\ba
\frac{1}{\vartheta_3^2\left(\pi x/a,e^{-\frac{\pi^2}{\alpha}}\right)}&=&1-4e^{-\frac{\pi^2}{\alpha}}\cos\left(\frac{2\pi}{a}x\right)+12e^{-\frac{2\pi^2}{\alpha}}\cos^2\left(\frac{2\pi}{a}x\right)-32e^{-\frac{3\pi^2}{\alpha}}\cos^3\left(\frac{2\pi}{a}x\right)
\nonumber \\
&+&e^{-\frac{4\pi^2}{\alpha}}\left[80\cos^4\left(\frac{2\pi}{a}x\right)-4\cos\left(\frac{4\pi}{a}x\right)\right]+{\cal O}\left(e^{-\frac{5\pi^2}{\alpha}}\right)\,,
\label{eq3-4-10}
\ea
}
from which we obtain:
\be
\frac{1}{a}\int_0^a{\rm d}x\frac{1}{\eta(x)}=\frac{1}{C_\alpha^{\prime 2}}\left[1+6e^{-\frac{2\pi^2}{\alpha}}+30e^{-\frac{4\pi^2}{\alpha}}+{\cal O}\left(e^{-\frac{5\pi^2}{\alpha}}\right)\right]\,.
\label{eq3-4-11}
\ee
Using Eqs.\,(\ref{eq2-13}), (\ref{eq3-2-1}), and (\ref{a-3}), we estimate:
\be
\frac{1}{C_\alpha^{\prime 2}}=\sqrt{\frac{\alpha}{2\pi}}I(\alpha)=1+2e^{-\frac{2\pi^2}{\alpha}}+{\cal O}\left(e^{-\frac{8\pi^2}{\alpha}}\right)\,,
\label{eq3-4-12}
\ee
and finally:
\be
f_s=1-8e^{-\frac{2\pi^2}{\overline{\alpha}}}+22e^{-\frac{4\pi^2}{\overline{\alpha}}}+\ldots
\label{eq3-4-13}
\ee
where the first term is linear in $\rho-\rho_c$ and the second is quadratic.

In the opposite limit of high densities,
\be
\psi(x)\simeq\frac{C_\alpha}{\sqrt{V}}e^{-\frac{\alpha}{a^2}x^2}\,,
\label{eq3-4-14}
\ee
and then
\be
f_s\ge 2e^{-\frac{\overline{\alpha}}{2}}>0\,.
\label{eq3-4-15}
\ee
In conclusion, according to both VT1 and VT2 the crystal is supersolid at all densities.

\section{Conclusions}

Pressure-driven clusterization of a fluid of soft-core bosons at $T=0$, i.e., the emergence of clumps of overlapping particles (clusters) under compression, is among the simplest examples of a quantum transition. At variance with two or three dimensions, where the formation of clusters is always accompanied by the appearance of crystalline order~\cite{Pomeau,Saccani,Prestipino3}, in 1D a no-go theorem by Pitaevskii and Stringari~\cite{Pitaevskii2} excludes the possibility of long-range order in the thermodynamic limit, implying loss of crystalline and phase coherence at large distances. However, (truncated) crystalline order is recovered in a large, but {\em finite}, 1D system confined in an elongated trap or placed in a narrow torus.

In this paper we apply three different MF theories to the study of clusterization in 1D, representing the ground state of the system as a pure condensate. At variance with 3D, where MF theory holds for small $\rho\epsilon$ values (with $\epsilon$ denoting the interaction strength), a 1D system approaches the weak-coupling regime for decreasing $\epsilon/\rho$. We further assume crystalline ordering of the dense phase. Besides GP theory, which is equivalent to selecting the best MF state, we consider two variational approximations for the single-particle wave function $\psi$: in one case, $\psi$ is written as a sum of Gaussians (a two-parameter ansatz); in the other case, by a similar sum we represent $\psi^2$. The virtue of variational theory is that the energy functional is written in almost closed form, which allows us to derive a number of analytic predictions.

In one dimension, the freezing transition turns out to be continuous and occurs at the highest density at which the fluid is still superfluid. As a rule, the crystalline ground state is a cluster crystal, meaning that the average site occupancy is larger than one. Moreover, the crystal is supersolid, meaning that the moment of inertia is smaller than for a classical solid of same mass and size. In more physical terms, supersolidity can be ascribed to the delocalization of the condensate wave function over the whole crystal and, particularly, to a non-zero probability of observing a particle in the interstitial region.

\appendix
\section{Some useful formulas}
\renewcommand{\theequation}{A.\arabic{equation}}

In this Appendix we collect a few formulas relative to Jacobi theta functions. These are special functions with a relation to elliptic functions. For a complex variable $z$ and a complex number $q$ of modulus less than 1, theta functions are defined as~\cite{jacobitheta}:
\ba
\vartheta_1(z,q)&\equiv&\sum_{n=-\infty}^{+\infty}(-1)^{n-1/2}q^{(n+1/2)^2}e^{(2n+1)iz}\,;
\nonumber \\
\vartheta_2(z,q)&\equiv&\sum_{n=-\infty}^{+\infty}q^{(n+1/2)^2}e^{(2n+1)iz}\,;
\nonumber \\
\vartheta_3(z,q)&\equiv&\sum_{n=-\infty}^{+\infty}q^{n^2}e^{2niz}\,;
\nonumber \\
\vartheta_4(z,q)&\equiv&\sum_{n=-\infty}^{+\infty}(-1)^nq^{n^2}e^{2niz}\,.
\label{a-1}
\ea
Clearly, a less symmetric form exists for each function, e.g.,
\be
\vartheta_3(z,q)=1+2\sum_{n=1}^\infty q^{n^2}\cos(2nz)\,.
\label{a-2}
\ee
For $z=0$, the $\vartheta_2$ and $\vartheta_3$ functions have the following obvious expansions around $q=0$:
\be
\vartheta_2(0,q)=2q^{1/4}(1+q^2+q^6+\ldots)\,;\,\,\,\vartheta_3(0,q)=1+2q+2q^4+2q^9+\ldots
\label{a-3}
\ee

Interestingly, there is a way to express the logarithmic derivative of $\vartheta_3$ in the form of a series~\cite{jacobitheta}:
\be
\frac{\vartheta_3'(z,q)}{\vartheta_3(z,q)}=4\sum_{n=1}^\infty(-1)^{n}\frac{q^{n}}{1-q^{2n}}\sin(2nz)\,.
\label{a-4}
\ee
A similar formula exists for each theta function.

A duality property holds for Jacobi theta functions~\cite{nistdigitallibrary}, by which a particular theta function for $q\lesssim 1$ is related to another theta function for $q\gtrsim 0$. For example,
\be
(-i\tau)^{1/2}\vartheta_2(z|\tau)=e^{i\tau'z^2/\pi}\vartheta_4(z\tau'|\tau')\,,
\label{a-5}
\ee
where, e.g., $\vartheta_2(z|\tau)$ stands for $\vartheta_2(z,q)$, with $q=\exp(i\pi\tau)$ and ${\rm Im}\,\tau>0$, whereas $\tau'=-1/\tau$. Another useful formula is:
\be
(-i\tau)^{1/2}\vartheta_3(z|\tau)=e^{i\tau'z^2/\pi}\vartheta_3(z\tau'|\tau')\,.
\label{a-6}
\ee


\begin{thebibliography}{99}
\bibitem{note} In equivalent terms, one might say that MF theory is valid whenever the kinetic energy per unit particle due to localization, $\hbar^2\rho^{2/d}/2m$, is much larger than the interaction energy $g\rho$, see e.g. \cite{Astrakharchik}.
\end{thebibliography}
\end{document}